\documentclass[preprint,aps,floats]{revtex4}
\usepackage{epsfig}

\usepackage{graphicx}% Include figure files
\usepackage{dcolumn}% Align table columns on decimal point
\usepackage{bm}% bold math
\def\lsim{\mathrel {\vcenter {\baselineskip 0pt \kern 0pt
    \hbox{$&lt;$} \kern 0pt \hbox{$\sim$} }}}
\def\gsim{\mathrel {\vcenter {\baselineskip 0pt \kern 0pt
    \hbox{$&gt;$} \kern 0pt \hbox{$\sim$} }}}

\newcommand{\U}{{\cal {U}}}
\begin{document}
%\preprint{hep-ph/0703270}

\title{Constraints on Unparticle Interactions from \\Invisible Decays of Z, Quarkonia and Neutrinos}

\author{Shao-Long Chen}
\email{shaolong@phys.ntu.edu.tw}
\author{Xiao-Gang He}
\email{hexg@phys.ntu.edu.tw}
\author{Ho-Chin Tsai}
\email{hctsai@phys.ntu.edu.tw} \affiliation{ Department of Physics
and Center for Theoretical Sciences, National Taiwan University,
Taipei, Taiwan}

\date{\today}

%abstract=================================================================
\begin{abstract}
Unparticles ($\U$) interact weakly with particles. The direct
signature of unparticles will be in the form of missing energy. We
study constraints on unparticle interactions using totally invisible
decay modes of $Z$, vector quarkonia $V$ and neutrinos. The
constraints on the unparticle interaction scale $\Lambda_\U$ are
very sensitive to the dimension $d_\U$ of the unparticles. From
invisible $Z$ and $V$ decays, we find that with $d_\U$ close to 1
for vector $\U$, the unparticle scale $\Lambda_\U$ can be more than
$10^4$ TeV, and for $d_\U$ around 2, the scale can be lower than one
TeV. From invisible neutrino decays, we find that if $d_\U$ is close
to 3/2, the scale can be more than the Planck mass, but with $d_\U$
around 2 the scale  can be as low as a few hundred GeV. We also
study the possibility of using $V (Z)\to \gamma + \U$ to constrain
unparticle interactions, and find that present data give weak
constraints.
\end{abstract}

%\pacs{PACS numbers: ***}

\maketitle

\noindent {\bf Introduction}

Recently Georgi proposed an interesting idea to describe possible
scale invariant effects at low energies by an operator $O_\U$,
termed unparticle~\cite{Georgi:2007ek}. Based on a specific scale
invariant theory with a non-trivial infrared fixed point by Banks
and Zaks~\cite{Banks:1982gt}, it was argued that operators
$O_{BZ}$ made of BZ fields may interact with operators $O_{SM}$
made of Standard Model (SM) fields at some high energy scale by
exchanging particles with large masses, $M_{\cal{U}}$, and induce
interactions of the form
\begin{eqnarray}
{\tilde C_{\cal{U}}\over M^{d_{SM}+d_{BZ} - 4}_{\cal{U}}}O_{SM}
O_{BZ}\;,\label{aa}
\end{eqnarray}
where $d_{BZ}$ and $d_{SM}$ are the dimensions of the operators
$O_{BZ}$ and $O_{SM}$.

At another scale $\Lambda_{\cal{U}}$ the BZ sector induces
dimensional transmutation, and below that scale the BZ operator
$O_{BZ}$ matches on to unparticle operator $O_{\cal{U}}$ with
dimension $d_\U$. The unparticle interaction with SM particles at
low energy then has the form
\begin{eqnarray}
C_\U \lambda \Lambda_{\cal{U}} ^{4-d_{SM} - d_\U} O_{SM}
O_{\cal{U}}\;,\;\;\;\; \lambda = \left ({\Lambda_{\cal{U}}\over
M_{\cal{U}}}\right )^{d_{SM}+d_{BZ}-4}\;.\label{bb}
\end{eqnarray}

The unparticle may have different Lorentz structures such as a
scalar $O_\U$, a vector $O_\U^{\mu}$, a spinor $O_\U^s$, and etc..
The specific form of SM particle and unparticle interactions are
not known and are usually parameterized in terms of operators. In
Ref.~\cite{Chen:2007qr} a class of operators involving SM particles
and unparticles are listed. Using these operators one can study
unparticle phenomenology in a systematic way.

One of the main phenomenological goals of unparticle physics study
is to find out at what energy scale unparticle effects may show
up [1, 3-27].
%%~\cite{Georgi:2007ek,Georgi:2007si,Cheung:2007ue,Luo:2007bq,Chen:2007vv,Ding:2007bm,
%%Liao:2007bx,Aliev:2007qw,Li:2007by,Lu:2007mx,Stephanov:2007ry,Fox:2007sy,Greiner:2007,
%%Davoudiasl:2007,Choudhury:2007,Chen:2007qr,Aliev:2007gr,Mathews:2007hr,Zhou:2007zq,
%%Ding:2007zw,Chen:2007je,Liao:2007ic,Bander:2007nd,Rizzo:2007xr,Cheung:2007ap,Goldberg:2007tt}.
The most direct signatures of unparticles will be in the form of
missing energy in invisible decays of particles with an unparticle
$\U$ in the final state. In this paper we study constraints on the
unparticle interactions using $Z\to \mbox{invisible}$, $V\to
\mbox{invisible}$, and $\nu \to \mbox{invisible}$ decays. We will
also study the possibility of using $V(Z)\to \gamma + \U$ to
constrain unparticle interactions.

\noindent
{\bf Constraints from invisible decay of $Z$ boson}

The process $Z\to \U$ contributes to invisible decay of $Z$. For
this process the following operators, with SM fields and
derivatives contribute less than or equal to 4 dimensions, will
contribute~\cite{Chen:2007qr}
\begin{eqnarray}
\lambda'_{bO}\Lambda_{\cal{U}}^{1-d_\U}B_{\mu\nu}\partial^\mu
O_\U^\nu\;, \;\;\;\;\tilde \lambda'_{bO}\Lambda_{\cal{U}}^{1-d_\U}\tilde
B_{\mu\nu}\partial^\mu O_\U^\nu\;,
\;\;\;\;\lambda'_{hh}\Lambda_{\cal{U}}^{1-d_\U}(H^\dagger D_\mu H)O^\mu_{\cal{U}}\;.
\end{eqnarray}
Here the vector unparticle operator $O_\U^\mu$ is hermitian and
transverse with, $\partial_\mu O^\mu_\U=0$.

The matrix elements for $Z \to \U^\mu$ resulting from the above
operators are given by~\cite{Chen:2007qr}
\begin{eqnarray}
&&M(Z \to \U^\mu, \lambda'_{bO}) = \lambda'_{bO}
\Lambda_{\cal{U}}^{1-d_{\cal{U}}} \sin\theta_W(k_Z\cdot k_O
\epsilon_Z\cdot \epsilon_O - k_Z\cdot
\epsilon_O k_O\cdot \epsilon_Z),\nonumber\\
&&M(Z \to \U^\mu, \tilde \lambda'_{bO}) = i\tilde \lambda'_{bO}
\Lambda_{\cal{U}}^{1-d_{\cal{U}}}
\sin\theta_W \epsilon_{\mu\nu\alpha\beta}k^\mu_Z \epsilon^\nu_Z k^\alpha_O \epsilon^\beta_O,\nonumber\\
&&M(Z\to \U^\mu, \lambda'_{hh}) = -2Im(\lambda'_{hh})
\Lambda_\U^{1-d_\U}{e\over \sin(2\theta_W)}{v^2\over 2}
\epsilon_Z\cdot \epsilon_O,
\end{eqnarray}
where $v=246$ GeV is the vacuum expectation value of the Higgs
doublet. Since $k_Z = k_O$ for $Z$ to $O^\mu_\U$ transition, the
second term in the above does not contribute. Here we have used
the notation $\epsilon_Z$ and $\epsilon_O$ to describe the
polarizations of $Z$ and $\U^\mu$.

For a decay of a particle into an unparticle and other particles, the differential decay rate is
given by~\cite{Georgi:2007ek}
\begin{eqnarray}
d\Gamma(P\to \U) = {|\overline{M}|^2  \over 2
m_P}d\Phi(P)\;,\label{general}
\end{eqnarray}
where $d\Phi(P)$ is the phase space factor for the decay. It is
given by
\begin{eqnarray}
d\Phi = \int (2\pi)^4 \delta^4 (P-\sum_j p_j) \prod_j d\Phi(p_j){d^4
p_j\over (2\pi)^4}.
\end{eqnarray}
For a particle the phase factor $d\Phi(p_j)$ is equal to $2\pi
\theta(p^0_j) \delta(p^2_j-m_j^2)$, and for an unparticle it is given by
$A_{d_\U} \theta(p^0)\theta(p^2) (p^2)^{d_\U -2}$ with $A_{d_\U}
=(16\pi^{5/2}/(2\pi)^{2d_\U})(\Gamma(d_\U+1/2)/\Gamma(d_\U -
1)\Gamma(2d_\U))$.

The decay width of a particle decay into an unparticle is given by
\begin{eqnarray}
\Gamma(P\to \U) &=& {|\overline{M}|^2\over 2 m_P} A_{d_\U}
(m^2_P)^{d_{\cal{U}}-2}\;.
\end{eqnarray}

We note that in the limiting case of $d_\U = 1$ the decay width becomes zero
since $A_{d_\U}$ has a factor $1/\Gamma(d-1)$ which goes to zero when $d_\U\to 1$.
Physically this is because that in this case~\cite{Georgi:2007ek}, $\lim_{d_\U \to 1^+} A_{d_\U} \theta(p^2)/p^{2(2-d_\U)}
= 2\pi \delta(p^2)$, the unparticle behaves as a massless particle. When $m^2_p \neq 0$, the
decay rate $\Gamma(P \to \U)$ is zero.

For $Z \to \U$, we have
\begin{eqnarray}
&&\Gamma(Z\to \U^\mu, \lambda'_{bO})={\Lambda_\U^2 \over 2 m_Z}
\sin^2\theta_W (\lambda'_{bO})^2 \left (\frac{m^2_Z}{
\Lambda_{\cal{U}}^2}\right )^{d_{\cal{U}}}
A_{d_\U}\;,\nonumber\\
&&\Gamma(Z\to \U^\mu, \lambda'_{hh})={\Lambda_\U^2 \over 2
m_Z}{4\pi\alpha\over \sin^2(2\theta_W)} (Im(\lambda'_{hh}))^2
\left ( {v^2\over m^2_Z}\right )^2 \left (\frac{m^2_Z}{
\Lambda_{\cal{U}}^2}\right )^{d_{\cal{U}}} A_{d_\U}\;.
\end{eqnarray}

In general four parameters are needed to describe unparticle
interactions with SM particles: $C_{\cal{U}}$, $\lambda$,
$\Lambda_{\cal{U}}$ and $d_{\cal{U}}$ as shown in eqs.(\ref{aa})
and (\ref{bb}). If $d_{SM} + d_{BZ}
>4 $, the parameters $\lambda$ is less than 1. The parameter
$C_\U$ contains information about the original heavy particle
mediating interaction of the SM and scale invariant sectors, and
also information about the transmutation. One may normalize the
parameter $C_\U$ into the definition of $\lambda$ for one
operator, but in general will not be able to do so for more than
one operator. The values for $C_\U$ depend on the detailed
dynamics. If one is only concerned with the scale where different
transitions have happened, one usually sets $C_\U$ to be one and
use the two parameters $\lambda$ and $\Lambda_\U$ to describe the
situation. In our numerical discussions, we will also follow this
subscription.

\begin{figure}[!htb]
\begin{center}
\includegraphics[width=10cm]{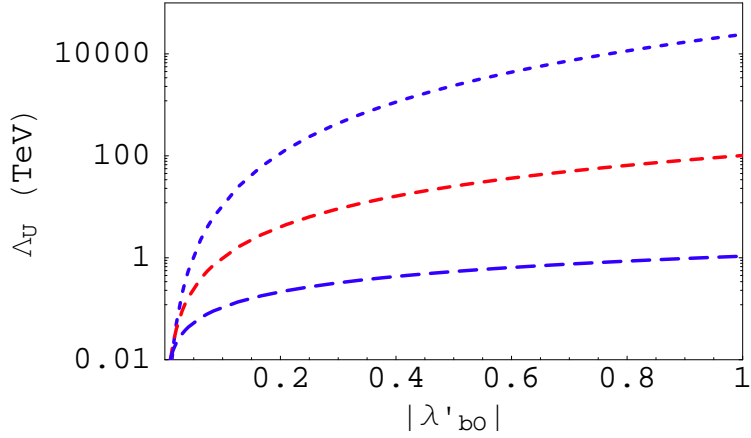}
\vspace*{-0.3cm}
\end{center}
\caption{Bounds on parameter space of $\Lambda_\U$ vs. $\lambda'_{bO}$
allowing unparticle decay mode to saturate the
difference of 3 MeV for invisible decay width of $Z$ between SM
prediction and experimental data. The lines from top to bottom are
for $d_\U=1.3, 1.5, 2.0$.} \label{diagram1}
\end{figure}

Precise experimental data have been obtained on Z decay
widths~\cite{Yao:2006px} with the invisible width to be:
$\Gamma(Z\to \mbox{invisible}) = 499.0\pm 1.5$ MeV. This is to be
compared with the width of $501.65\pm 0.11$ MeV from SM prediction
for $Z$ decy into neutrinos. New contribution to invisible $Z$
decay is therefore constrained severely, basically need to be
within the range of experimental error bar of order one MeV. In
Fig.~\ref{diagram1} we show constraints on unparticle interactions
allowing the unparticle contribution to invisible $Z$ decay to
saturate 2$\sigma$ error of experimental data of 3 MeV.
Numerically the bound on $Im(\lambda'_{hh})$ is about $5.5$ times
stronger than $\lambda'_{bO}$ for given $d_\U$ and $\Lambda_\U$.
In Fig.~\ref{diagram1} we only show constraint on $\lambda'_{bO}$.
Since for a given $d_\U$, the scale $\Lambda_\U$ also depends on
the parameter $\lambda'_{bO}$, one can view the constraints on
$\lambda'_{bO}$ for a given $\Lambda_\U$ or on the scale
$\Lambda_\U$ for a given $\lambda'_{bO}$. In any case, from
Fig.\ref{diagram1} it is clear that the constraints are very
sensitive to the dimension parameter $d_\U$. If $d_\U$ is close to
1, for example with $\lambda'_{bO} = 1$ and $d_\U$ to be 1.3,
$\Lambda_\U$ needs to be larger than $10^4$ TeV, but $\Lambda_\U$
can be as low as one TeV for $d_\U = 2$. If by some
means the scale $\Lambda_\U$ is known, for example $\Lambda_\U=1$
TeV, we have the upper bounds $\lambda'_{bO}=0.049$ and $0.10$ for
the cases $d_\U$=$1.3$ and $1.5$, respectively.

\noindent
{\bf Constraints from invisible decay of quarkonia}

For a vector quarkonium $V$ decays into an unparticle $\U$ the
following operators will contribute.
\begin{eqnarray}
&&\lambda'_{QQ}\Lambda_{\cal{U}}^{1-d_\U}\bar Q_L \gamma_\mu Q_L
O^\mu_{\cal{U}}, \;\lambda'_{UU}\Lambda_{\cal{U}}^{1-d_\U}\bar U_R
\gamma_\mu U_R O_{\cal{U}}^\mu,
\;\lambda'_{DD}\Lambda_{\cal{U}}^{1-d_\U}\bar D_R  \gamma_\mu D_R
O^\mu_{\cal{U}}.
\end{eqnarray}

The matrix elements for vector
quarkonia and unparticle transition resulting from the above interactions can be written
as the following
\begin{eqnarray}
M(V\to \U^\mu, \lambda') = {1\over 2} \lambda' \Lambda_\U^{1-d_\U}
\langle 0|\bar q \gamma_\mu q |V\rangle \cdot \epsilon_O^\mu,
\end{eqnarray}
where for $q$ being an up type quark, $\lambda'$ can be
$\lambda'_{QQ}$ and $\lambda'_{UU}$ with $Q_q = 2/3$, and for $q$
being a down type quark, $\lambda'$ can be $\lambda'_{QQ}$ and
$\lambda'_{DD}$ with $Q_q= -1/3$.

We obtain
\begin{eqnarray}
\frac{Br(V \to\U^\mu, \lambda')}{Br(V \to\mu^+\mu^-)}=
\frac{3A_{d_\U}|\lambda'|^2} {32\pi \alpha^2 Q_q^2}
\left(\frac{m_V^2}{\Lambda_\U^2}\right)^{d_\U-1}\;.
\end{eqnarray}

The operator
$\lambda'_{bO}\Lambda_{\cal{U}}^{1-d_\U}B_{\mu\nu}\partial^\mu
O^\nu$ also contributes to this process. We have
\begin{eqnarray}
\frac{Br(V \to\U^\mu, \lambda'_{bO})}{Br(V \to\mu^+\mu^-)}=
\frac{3A_{d_\U}\cos^2\theta_W|\lambda'_{bO}|^2} {2\alpha}
\left(\frac{m_V^2}{\Lambda_\U^2}\right)^{d_\U-1}\;.
\end{eqnarray}

\begin{figure}[!htb]
\begin{center}
\includegraphics[width=10cm]{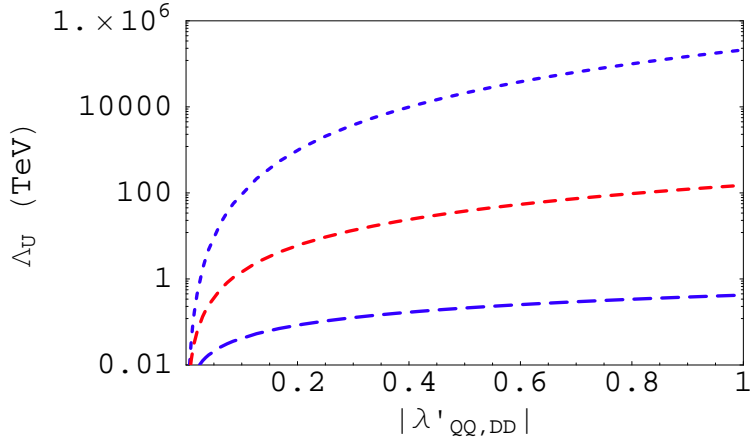}
\vspace*{-0.3cm}
\end{center}
\caption{Bounds on parameter space of $\Lambda_\U$ vs.
$\lambda'_{QQ,DD}$ allowing unparticle decay mode to saturate the
experimental data for invisible $\Upsilon$ decay. The lines from
top to bottom are for $d_\U=1.3, 1.5, 2.0$.} \label{diagram2}
\end{figure}

In Fig.~\ref{diagram2} we show constraints on unparticle
interactions using experimental data~\cite{Tajima:2006nc}:
$Br(\Upsilon\to \mbox{invisible}) < 2.5 \times 10^{-3}$. For this
case $Q=D=b$ and $Q_q = -1/3$. In obtaining the constraints, we
have neglected small contributions from $\Upsilon \to \nu \bar
\nu$ to $\Upsilon$ invisible decay width. In this case, for given
$d_\U$ and $\Lambda_\U$, the constraints on $\lambda'_{QQ}$ and
$\lambda'_{DD}$ are the same, while the constraint on $\lambda'_{bO}$
is $5.6$ times weaker than the bounds for $\lambda'_{QQ, DD}$. In
Fig.~\ref{diagram2}, we show constraints on $\lambda'_{QQ,DD}$ and
$\Lambda_\U$. Again we see that the bounds are very sensitive to
the dimension $d_\U$. If $d_\U$ is close to 1, with $\lambda'_{QQ}
= 1$ and $d_\U = 1.3$, $\Lambda_\U$ needs to be larger than $2
\times 10^5$ TeV, but $\Lambda_\U$ can be as low as $400$ GeV for
$d_\U = 2$. If the scale $\Lambda_\U$ is set to be $1$ TeV, we
find the upper bounds $\lambda'_{QQ,DD}=0.025$ and $0.082$ for the
cases $d_\U$=$1.3$ and $1.5$ respectively. Note that the
constraint on $\lambda'_{bO}$ obtained from $\Upsilon \to \U$ is
weaker than that obtained from $Z\to \U$ by a factor of
$1.49({m_Z}/{m_\Upsilon})^{d_\U-1}$, which is in the range
$2.9\sim 14.4$ when $d_\U$ is in the range of $1.3\sim 2.0$.

One can easily work out the case for invisible decay of $J/\psi$
by taking $Q = c$, $U = c$ and $Q_q = 2/3$. BES has accumulated
more than a million $J/\psi$, it would be interesting to see if
these data when analyzed for invisible decay, a better constraint
could be obtained.

\noindent
{\bf Constraints from invisible decay of neutrinos}

For an active neutrino decays into an unparticle, the following operator
will contribute
\begin{eqnarray}
\lambda_s\Lambda_{\cal{U}}^{3/2-d_\U}\bar L_L H O^s_{\cal{U}}\;.
\end{eqnarray}

After the Higgs develops its vev, one obtains a transition matrix element between a neutrino and
an unparticle
\begin{eqnarray}
M(\nu \to \U^s) = \lambda_s \Lambda_{\cal{U}}^{3/2-d_\U}\bar \nu_L
{v\over \sqrt{2}} O^s_{\cal{U}}.
\end{eqnarray}
This leads to
\begin{eqnarray}
\Gamma(\nu \to \U^s) = {1\over 4} |\lambda_s|^2 {v^2\over m_\nu}
\left ( {m^2_\nu \over \Lambda^2_{\cal{U}}}\right )^{d_{\cal{U}} -
3/2}.
\end{eqnarray}

\begin{figure}[!htb]
\begin{center}
\includegraphics[width=10cm]{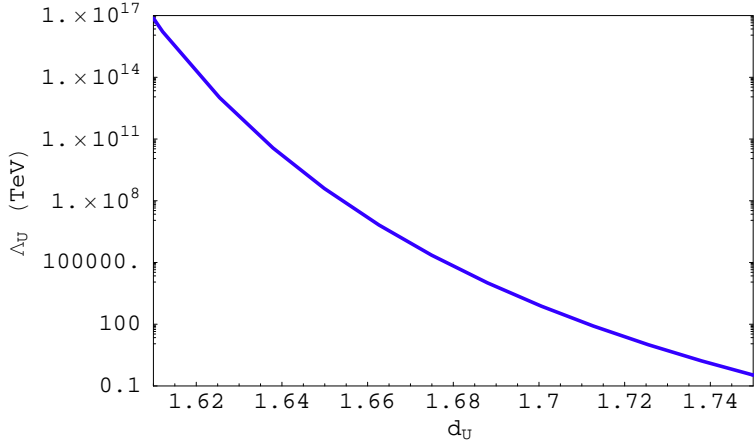}
\vspace*{-0.3cm}
\end{center}
\caption{Allowed parameter space for $\Lambda_{\cal{U}}$ vs.
$d_\U$.} \label{diagram3}
\end{figure}

Using constraints from solor neutrino data~\cite{Beacom:2002vi}
${\tau/ m } > 10^{-4}{\mbox{s}/ \mbox{eV}}$ on neutrino lifetime
and mass ratio, one can obtain information about the unparticle
interactions. Since the term generating this invisible neutrino
decay is related to the Yukawa coupling, it is natural to have
$\lambda_s v/\sqrt{2}$ to be of order the neutrino mass itself
(one can easily converts into different normalization). In this
case one would obtain
\begin{eqnarray}
m^2_\nu \left ({m^2_\nu\over \Lambda_\U^2}\right )^{d_\U - 3/2} <
1.3 \times 10^{-11} \mbox{eV}^2.
\end{eqnarray}
Applying this formula to the solar neutrino with the constraint on
its relevant mass to be larger than $\sqrt{\Delta m^2_{solar}}$,
we obtain the bound on $\Lambda_\U$ as a function of $d_\U$ in
Fig.~\ref{diagram3} with the central value of $\Delta m^2_{solar} =
8.0\times 10^{-5}$ eV$^2$~\cite{Yao:2006px}. It can be easily seen
that the bound on $\Lambda_\U$ depends on $d_\U$ very sensitively.
If one sets $\Lambda_\U$ to be less than the Planck scale $m_P
= 1.22\times 10^{19}$ GeV, the dimension $d_\U$ must be bigger than $1.6$.

\noindent {\bf Constraints from radiative $V\to \gamma
 + \mbox{invisible}$}

Now we study the possibility of using $V (Z)\to \gamma + \U$ to
constrain the unparticle interactions. Using the general formula
in eq.(\ref{general}), we obtain the differential rate for $V\to
\gamma + \U$ for a given matrix element $M$,
\begin{eqnarray}
\frac{d\Gamma(P\to \gamma + \U)}{dE_\gamma} &=&
\frac{|\overline{M}|^2}{2 m_P} A_{du}
(P^2_\U)^{d_\U-2}\frac{E_\gamma}{4 \pi^2}\;.
\end{eqnarray}

For $Z (V) \to \U$ process, the unparticle $\U$ must be a vector
type. For $Z(V) \to \gamma + \U$, the unparticle can be a scalar
or a vector. Using this process, constraints on scalar unparticle
interactions can also be obtained. We find that the following
operators contribute to $V \to \gamma + \U$ at the tree level,
\begin{eqnarray}
a) &&\lambda_{ww}
\Lambda_{\cal{U}}^{-d_\U}W^{\mu\nu}W_{\mu\nu}O_{\cal{U}},
\;\lambda_{bb}\Lambda^{-d_\U}_{\cal{U}}B^{\mu\nu}B_{\mu\nu}O_{\cal{U}},\;\tilde
\lambda_{ww} \Lambda_{\cal{U}}^{-d_\U}\tilde
W^{\mu\nu}W_{\mu\nu}O_{\cal{U}},
\;\tilde \lambda_{bb}\Lambda^{-d_\U}_{\cal{U}}\tilde B^{\mu\nu}B_{\mu\nu}O_{\cal{U}},\nonumber\\
b) &&\lambda_{QQ}\Lambda_{\cal{U}}^{-d_\U} \bar Q_L \gamma_\mu
D^\mu Q_L O_{\cal{U}}, \;\lambda_{UU}\Lambda_\U^{-d_\U}\bar
U_R\gamma_\mu D^\mu U_R O_{\cal{U}},
\;\lambda_{DD}\Lambda_{\cal{U}}^{-d_\U}\bar D_R \gamma_\mu D^\mu D_R O_{\cal{U}},\nonumber\\
c) &&\tilde \lambda_{QQ}\Lambda_{\cal{U}}^{-d_\U}\bar Q_L
\gamma_\mu Q_L
\partial^\mu O_{\cal{U}}, \;\tilde \lambda_{UU}\Lambda_{\cal{U}}^{-d_\U}\bar U_R \gamma_\mu
U_R \partial^\mu O_{\cal{U}},
\;\tilde \lambda_{DD}\Lambda_{\cal{U}}^{-d_\U}\bar D_R  \gamma_\mu D_R \partial^\mu O_{\cal{U}},\\
d) &&\lambda'_{QQ}\Lambda_{\cal{U}}^{1-d_\U}\bar Q_L \gamma_\mu
Q_L O^\mu_{\cal{U}}, \;\lambda'_{UU}\Lambda_{\cal{U}}^{1-d_\U}\bar
U_R \gamma_\mu U_R O_{\cal{U}}^\mu,
\;\lambda'_{DD}\Lambda_{\cal{U}}^{1-d_\U}\bar D_R  \gamma_\mu D_R
O^\mu_{\cal{U}},\;\lambda'_{bO}\Lambda_{\cal{U}}^{1-d_\U}B_{\mu\nu}\partial^\mu
O^\nu\;.\nonumber
\end{eqnarray}

For class a) contributions, we obtain
\begin{eqnarray}
\frac{Br(V\to \gamma + \U,\lambda_{GG})}{Br(V\to\mu^+\mu^-)}=\int
dE_\gamma
\frac{A_{d_\U}E_\gamma^3(\lambda_{GG})^2}{\pi^2\alpha(\Lambda_\U^2)^{d_\U}
(m_V^2-2m_VE_\gamma)^{2-d_\U}}\;,
\end{eqnarray}
where $\lambda_{GG}$ takes the values $\lambda_{ww}\sin^2\theta_W$
, $\lambda_{bb}\cos^2\theta_W, \tilde\lambda_{ww}\sin^2\theta_W$
and $\tilde\lambda_{bb}\cos^2\theta_W$ for the four operators in
class a) in order, respectively.

For classes b) and c) contributions, we have
\begin{eqnarray}
&&\frac{Br(V\to \gamma + \U,\lambda)}{Br(V\to\mu^+\mu^-)}=\int
dE_\gamma
\frac{A_{d_\U}m_V^2E_\gamma\lambda^2}{8\pi^2\alpha(\Lambda_\U^2)^{d_\U}
(m_V^2-2m_VE_\gamma)^{2-d_\U}}\;,\nonumber\\
&&\frac{Br(V\to \gamma + \U,\tilde
\lambda)}{Br(V\to\mu^+\mu^-)}=\int dE_\gamma
\frac{A_{d_\U}m_V^2E_\gamma\tilde\lambda^2}{4\pi^2\alpha(\Lambda_\U^2)^{d_\U}
(m_V^2-2m_VE_\gamma)^{2-d_\U}}\;.
\end{eqnarray}

For class d), contributions from the first three operators are
give by
\begin{eqnarray}
\frac{Br(V\to \gamma + \U,\lambda')}{Br(V\to\mu^+\mu^-)}=\int
dE_\gamma
\frac{A_{d_\U}(m_V^2-m_VE_\gamma)E_\gamma\lambda'^2}{2\pi^2\alpha(\Lambda_\U^2)^{d_\U-1}
(m_V^2-2m_VE_\gamma)^{3-d_\U}}\;.\label{vgu}
\end{eqnarray}

In the above $\lambda =\lambda_{QQ}$ or $\lambda_{DD}$ and
$\lambda$ = $\lambda_{QQ}$ or $\lambda_{DD}$ for quarkonia
composed of down and up type of quarks, respectively. Similarly
for $\tilde \lambda$ and $\lambda'$.

The fourth operator
$\lambda'_{bO}\Lambda_{\cal{U}}^{1-d_\U}B_{\mu\nu}\partial^\mu
O^\nu$ in class d) also contributes to $V\to \gamma + \U$ and the
decay width can be obtained by replacing $\lambda'$ with
$\lambda'_{bO}(eQ_q \cos\theta_W)$ in eq.(\ref{vgu}).

It is interesting to note that for contributions from classes a),
b) and c), $d_\U$ needs to be larger than 1 in order to have a
finite width for $V\to \gamma + \U$, while for the contributions
from class d), $d_\U$ needs to be larger than 2 to have a finite
width. In our numerical analysis, we will let $d_\U$ to be larger
than 2 for this case. Also note that the distributions of
$E_\gamma$ for class a), classes b) and c), and class d) are
different. This can be used to distinguish different contributions
if enough data are accumulated.

\begin{figure}[!htb]
\begin{center}
\includegraphics[width=7cm]{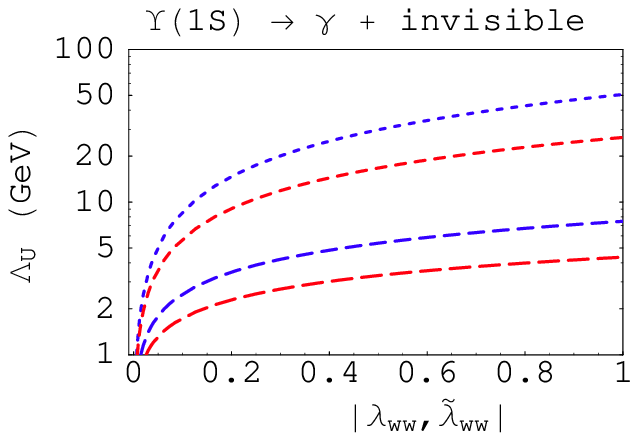}
\includegraphics[width=7cm]{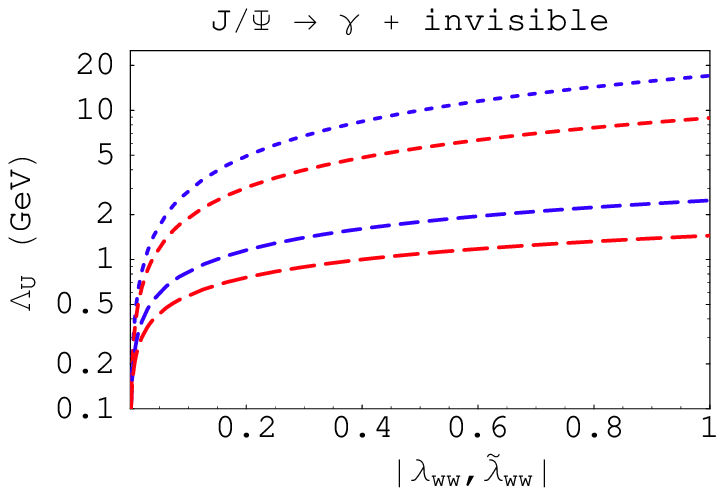}
\vspace*{-0.1cm}
\includegraphics[width=7cm]{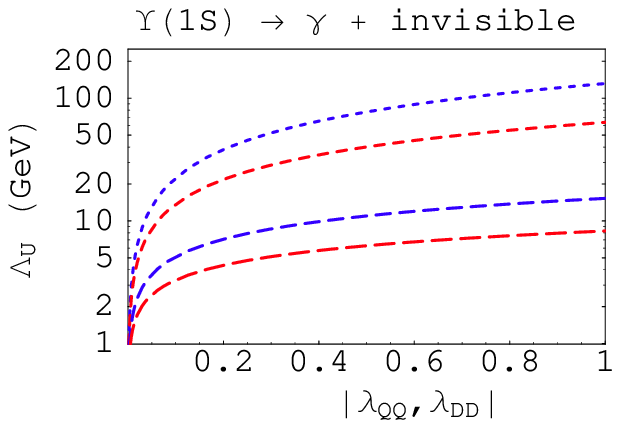}
\includegraphics[width=7cm]{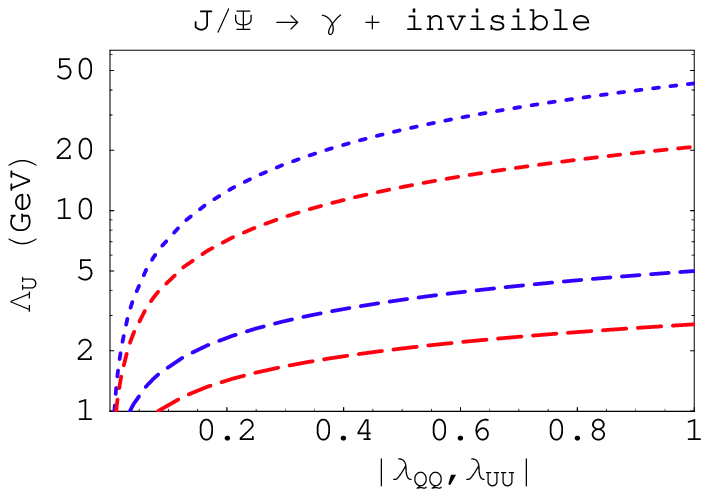}
\vspace*{-0.1cm}
\includegraphics[width=7cm]{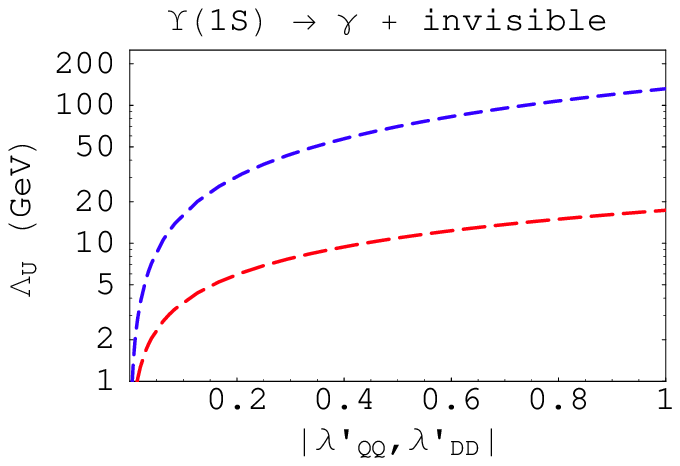}
\includegraphics[width=7cm]{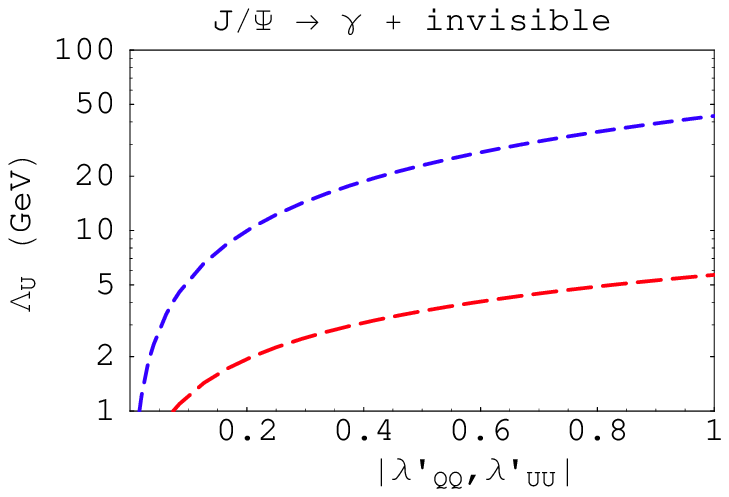}
\vspace*{-0.1cm}
\end{center}
\caption{Bounds on parameter space of $\lambda$ vs. $\Lambda_\U$
allowing unparticle decay mode to saturate the experimental data
for $\gamma+ \mbox{invisible}$ of $\Upsilon$ and $J/\psi$ decays.
From top to bottom, the lines represent the cases for $d_\U =1.3,
1.5, 2.1, 2.5$ and for the bottom panel, $d_\U$ only takes $2.1$
and $2.5$~. } \label{diagram4}
\end{figure}

There are experimental constrains on $\gamma + \mbox{invisible}$
decays of $\Upsilon$ and $J/\psi$ with $Br(J/\psi \to \gamma +
\mbox{invisible}) < 1.4\times 10^{-5}$~\cite{Edwards:1982zn}, and
$Br(\Upsilon(1S) \to \gamma + \mbox{invisible}) < 1.5\times
10^{-5}$~\cite{Antreasyan:1990cf,balest}. Combining the formula
obtained above for unparticle contributions, one can set
constraints on unparticle interactions. The results are shown in
Fig.~\ref{diagram4}. We find that the present upper bounds on
$\Upsilon (J/\psi) \to \gamma + \mbox{invisible}$ decay widths do
not give strong constraints on the unparticle interactions.
Improved bounds can provide more information.

We comment that there are also contributions to {$Z\to\gamma +
\U$} from class a) operators. The decay width is given by
\begin{eqnarray}
\frac{d\Gamma}{dE_\gamma}=\frac{A_{d_\U}\sin^2(2\theta_w)}{3\pi^2}\frac{E_\gamma^3m_Z}
{(m_Z^2-2m_ZE_\gamma)^2}\left(\frac{m_Z^2-2m_ZE_\gamma}{\Lambda_\U^2}\right)^{d_\U}
\lambda^2\;,
\end{eqnarray}
where $\lambda$ can be any of $\lambda_{ww,bb}, \tilde\lambda_{ww,bb}$.
It would be interesting to see if strong constraints can be
obtained for unparticle interactions when LEP data are analyzed
for $Z\to \gamma + \mbox{invisible}$.

\noindent {\bf Summary}

If unparticles exist they must interact weakly with particles. The
direct signature of unparticles will be in the form of missing
energy in decays and collisions of particles. In this paper we
have studied constraints on unparticle interactions using totally
invisible decay modes of $Z$, vector quarkonia $V$ and neutrinos.
There are several operators which can contribute to these decays.

Two operators with couplings $\lambda'_{hh}$ and $\lambda'_{bO}$
contribute to $Z\to \U$. Numerically the bound on
$Im(\lambda'_{hh})$ is about $5.5$ times stronger than
$\lambda'_{bO}$ for given $d_\U$ and $\Lambda_\U$. The constraints
are very sensitive to the dimension parameter $d_\U$. If $d_\U$ is
close to 1, with $\lambda'_{bO} = 1$ and $d_\U = 1.3$,
$\Lambda_\U$ needs to be larger than $10^4$ TeV, but $\Lambda_\U$
can be as low as one TeV for $d_\U = 2$. If by some
means that the scale $\Lambda_\U$ is known, for example
$\Lambda_\U=1$ TeV, we have the upper bounds $\lambda'_{bO}=0.049$
and $0.10$ for the cases $d_\U$=$1.3$ and $1.5$, respectively.

Several operators contribute to $V\to \U$, including the operator
with coupling $\lambda'_{bO}$ and additional ones
$\lambda'_{QQ,UU,DD}$. There is experimental upper bound from
$\Upsilon \to \mbox{invisible}$. We find that the constraints on
$\lambda'_{bO}$ are weaker than the constraint on
$\lambda'_{QQ,DD}$ by a factor of $5.6$.
The constraints are again sensitive to $d_\U$. If $d_\U$ is close
to 1, with $\lambda'_{QQ} = 1$ and $d_\U = 1.3$, $\Lambda_\U$
needs to be larger than $2\times 10^5$ TeV, but $\Lambda_\U$ can
be as low as $400$ GeV for $d_\U = 2$. If the scale $\Lambda_\U$
is set to be 1 TeV, we find the upper bounds
$\lambda'_{QQ,DD}=0.025$ and $0.082$ for the cases $d_\U$=$1.3$
and $1.5$ respectively. The constraint on $\lambda'_{bO}$ obtained
from $\Upsilon \to \U$ is weaker than that obtained from $Z\to \U$
by a factor of $1.49({m_Z}/{m_\Upsilon})^{d_\U-1}$, which is in
the range $2.9\sim 14.4$ when $d_\U$ takes value in the range of
$1.3\sim 2.0$.

There is one operator which can induce neutrino to $\U$ decay.
Strong constraint could be obtained using constraint on $\tau/m$
obtained from solar neutrino data. If one sets $\Lambda_\U$ to be
less than the Planck scale $m_P = 1.22\times 10^{19}$
GeV, the dimension $d_\U$ must be bigger than $1.6$.

We also studied the possibility of using $V (Z)\to \gamma + \U$ to
constrain unparticle interactions. We find that present
experimental upper bounds for $\Upsilon(J/\psi) \to \gamma +
\mbox{invisible}$ does not give strong bounds on unparticle
interactions.

\vskip 1.0cm \noindent {\bf Acknowledgments}$\,$ We thank
Chuan-Hung Chen and Cheng-Wei Chiang for bringing
references~\cite{Edwards:1982zn,Antreasyan:1990cf,balest} to our
attention. The work of authors was supported in part by the NSC
and NCTS.

\end{document}